# All twist and no bend makes raft edges splay: Spontaneous curvature of domain edges in colloidal membranes

Joia M. Miller[1, 4], Doug Hall[2], Joanna Robaszewski[1, 4], Prerna Sharma[3], Michael F. Hagan[1], Gregory M. Grason[2], Zvonimir Dogic[1,4]

[1]*Department of Physics, Brandeis University, Waltham, MA 02454, USA*
[2]*Department of Polymer Science and Engineering, University of Massachusetts, Amherst, MA 01003, USA*
[3]*Department of Physics, Indian Institute of Science, Bangalore 560012, India*
[4]*Department of Physics, University of California, Santa Barbara, CA 93106, USA*

**Abstract:** Using a combination of theory and experiments we study the interface between two immiscible domains in a colloidal membrane composed of rigid rods of different lengths. Geometric considerations of rigid rod packing imply that a domain of sufficiently short rods in a background membrane of long rods is more susceptible to twist than the inverse structure, a long-rod domain in a short-rod membrane background. The tilt at the inter-domain edge forces splay, which in turn manifests as a spontaneous edge curvature whose energetics are controlled by the length asymmetry of constituent rods. A thermodynamic model of such tilt-curvature coupling at inter-domain edges explains a number of experimental observations, including a non-monotonic dependence of the edge twist on the domain radius, and annularly shaped domains of long rods. Our work shows how coupling between orientational and compositional degrees of freedom in two-dimensional fluids give rise to complex shapes and thermodynamics of domains, analogous to shape transitions in 3D fluid vesicles.

**Popular Summary**: Immiscible fluids, such as oil and water, bulk phase separate leading to formation of liquid droplets. In 3D isotropic fluid mixtures droplets assume, on average, a spherical shape which minimizes the interfacial area, and thus the system free energy. Designing reconfigurable liquid droplets that adopt a broader class of non-spherical shapes remains an experimental challenge with on robust solution.

We study liquid-liquid phase separation in a binary colloidal membrane composed of length asymmetric achiral rods, demonstrating that the behavior of membrane-imbedded 2D liquid droplets is fundamentally different from their 3D bulk counterparts. Using a combination of theory and experiments we describe a universal geometric mechanism which yields an interfacial tension between two liquids domains that selects a preferred edge curvature. Curvature-dependent interfacial tension has important consequences on the structure of membrane imbedded droplets. It selects stable finite-sized droplets independent of the constituent's chirality. It also destabilizes large circular droplets, giving rise to an array of unique non-convex and annular domain shapes.

The consequence of spontaneous curvature has been studied for 3D lipid bilayer vesicles, where it generates a spectrum of complex and non-convex shapes. Length-asymmetric, binary rod membranes constitute 2D analogs to these 3D structure. Membrane embedded 2D droplets also demonstrate a robust pathway for assembly of reconfigurable shape-changing colloidal structures, which emerge from collective assembly of simple rodlike building blocks. More broadly, the symmetry-based principles underlying these phenomena may be relevant for phase-separated lipid bilayer rafts, an important self-organizing principle of biological cells which are molecular analogs of micron-sized colloidal membranes.

**Introduction:** A membrane's local composition is geometrically coupled to its structure. As such, domains of different composition deform the local membrane structure, generating interactions and instabilities that have no analogs in 3D isotropic solvents (*1-5*). For example, domains can locally change the preferred membrane thickness or curvature, and such deformations propagate into the membrane interior, generating effective long-ranged interactions (*6*). In particular, two component membranes exhibit a pronounced tendency to demix into distinct domains that are separated by a 1D interface (*7-11*). In biology, such phase-separated finite-sized domains, called lipid rafts, acquire important functional roles (*12-15*). In synthetic monolayers and bilayer membranes, phase separated domains are ubiquitous and can assume intriguing non-circular shapes (*16-18*). Notwithstanding significant progress, the interplay between orientational and compositional order in fluid membranes and its effect on domain size, morphologies and thermodynamics remain poorly understood.

Colloidal membranes are one rod-length thick liquid-like monolayer that spontaneously assemble from a mixture of monodisperse colloidal rod-like particles and non-adsorbing depleting polymers (*19-22*). They are micron sized analogues of conventional lipid-bilayers that provide a unique opportunity to visualize and quantify membrane inclusions and other superstructures and elucidate the fundamental laws that govern their behaviors. We study the structure of inter-domain edges in a phase-separated colloidal membrane that spontaneously assemble in a mixture of bidisperse rod-like colloids and non-adsorbing polymer. A geometric argument predicts that edge curvature towards shorter rod domains softens the resistance of the edge to twist, which in turn gives rise to a spontaneous non-zero edge curvature. A theoretical model that captures the twist-curvature coupling of the interface is supported by experimental observations that include (i) enhancement or suppression of edge tilt for respectively the short-rod and long-rod rafts, (ii) a non-monotonic dependence of the edge twist on the geodesic curvature of the edge, and (iii) the shape instability of the circular long-rod rafts, which above a critical size spontaneously transform into annular liquid droplet. Previous work showed that chiral rod-like inclusions in colloidal membranes experience long-ranged repulsive interactions that stabilize curved interfaces and assembly of finite-sized rafts (*23-25*). Our present findings demonstrate a chirality-independent mechanism that drives spontaneous interfacial twist in membranes composed of rods with sufficiently different lengths, due to coupling of the twist and curvature to overall membrane volume. The same mechanism could stabilize assembly of finite-sized colloidal rafts even in weakly chiral or achiral systems (*26*). More broadly, these results show that the interplay between orientation and composition in phase-separated membranes gives rise to shape selection mechanisms of 2D droplets that are reminiscent of those studied in 3D vesicles (*27*).

**Geometry of length-asymmetric rod domains:** We consider a phase-separated membrane composed of rigid, rod-like particles with two different lengths, $\ell_{in} \neq \ell_{out}$. We assume that all rods have smectic-like order with their centers confined to the 2D mid-plane. The phase separation occurs by formation of a circular raft of radius $R$ composed of rods of length $\ell_{in}$, surrounded by a background membrane composed of the alternate length rods, $\ell_{out}$. The local rod alignment is described by an axisymmetric double-twist whirlpool-like pattern of tilt (**Fig 1a, c**). In polar coordinates, such director field around the raft center can be specified in terms of mid-plane position $(r_0, \phi)$ of the rod centers as $\mathbf{n}(r_0, \phi) = \cos\theta(r_0)\,\hat{z} + \sin\theta(r_0)\,\hat{\phi}$, where $\theta(r_0)$ is the tilt angle relative to the membrane normal, $\hat{z}$. We consider profiles where the director at the raft center aligns with the membrane normal ($\theta(0) = 0$), twists to a finite value at the raft edge ($\theta(r_0 = R) = \theta_0$), and then unwinds back to the membrane normal far away from the raft ($\theta(r_0 \to \infty) = 0$). Previous models have shown that such a director pattern can be driven by asymmetric preferences for cholesteric twist in the distinct domains (*23, 24*). Here, we describe a chirality-independent mechanism in which the asymmetry between two rod lengths renders the domain interface highly susceptible to tilt at curved edges. The radial twist pattern leads to a generic coupling between orientation twist and density (**Fig. 1b**). When the rod's persistence length is much larger than the its contour length, bend is expelled from the raft, and the rod trajectories at radial position, $r_0$, can be described by $\mathbf{r}(r_0, \phi, z_0) = r_0 \hat{r} + z_0 \mathbf{n}$, where $z_0 \in [-\ell/2, \ell/2]$ is the arc-length along the rod relative to the mid-plane. Hence, surfaces of fixed $r_0$ are *ruled surfaces*, spanned by a set of straight lines (i.e. the rod trajectories) which maintain constant tilt $\theta(r_0)$ (**Fig. 1b**). The radial distance of rods from the raft center is: $r_\perp(r_0, z) = \sqrt{r_0^2 + z^2 \tan^2 \theta(r_0)}$, where $z = z_0 \cos \theta(r_0)$ is the raft vertical height. Consequently, rods at fixed $r_0$ lie in a *hyperboloid,* implying that the raft geometry can be decomposed into nested hyperboloids whose geometry is specified by $r_0$ and $\theta(r_0)$.

Splayed liquid crystalline configurations, where $\nabla \cdot \mathbf{n} \neq 0$, are costly because they introduce variation away from the preferred inter-rod (areal) density $\rho$. This follows from the "conservation law" of Meyer and deGennes: $(\mathbf{n} \cdot \nabla)\rho = -\rho(\nabla \cdot \mathbf{n})$, which holds for membranes in which rod-ends are expelled to the top and bottom boundary (*28, 29*). Along the mid-plane ($z_0 = 0$), the raft director field is given by the splay-free double twist pattern $\mathbf{n}(r_0, \phi)$. However, the combination of zero bend and variable tilt introduces splay away from the mid-plane, forcing rods to deviate from constant spacing (*30*). Graphically, this can be seen from the "frayed" geometry of the nested hyperboloids as the rods tilt away from $\hat{z}$ (**Fig. 1b**). Mathematically, this splay can be related to radial tilting of the hyperboloids $\frac{\partial r_\perp}{\partial z} \cong z \tan^2 \theta(r_\perp)/r_\perp$, where $r_\perp \cong r_0$ and we consider small $z$ for simplicity. Because rods live on these surfaces, the radial component of the director is $\frac{\partial r_\perp}{\partial z}$. Near the mid-surface $\nabla \cdot \mathbf{n} = r^{-1} \partial_r (r\, n_r) \propto z\, \partial_r(\tan^2 \theta)/r$ (Appendix B eq. B9). Consequently, splay arises away from the membrane midplane due to the *radial gradients* of tilt. In turn, the conservation law implies that with increasing splay the rod density $\rho$ drops away from the mid-plane. This local density expansion occurs at the top and the bottom of the inner raft edge. Furthermore, a decreasing tilt profile in the outer membrane, where rods tilt back towards the membrane normal, generates compression, with the local density being highest at the rod ends (**Fig. 1e, f**). The expansion and compression of the inner and outer raft edges is a consequence of the variable double-twist texture.

The splay-twist coupling has a significant consequence on the structure and thermodynamics of the domain interface, that has not been considered previously (*23, 24*). We first consider a model of colloidal monolayers of achiral rod-like molecules that are condensed by the depletant polymers. Entropy favors the maximal volume accessible to the depletants, which occurs for membrane configurations that minimize volume, since the membrane is inaccessible to the depletant (*21*). Note that a straight, zero-curvature interface between long and short rods would have a square-like cross-sectional profile, and rod tilt parallel to the interface would not introduce splay (**Fig. 2c, Movie S1**). Curving such an interface towards short rods produces a double twisted director field upon tilting of the rods. In turn, the geometric argument outlined above illustrates that this causes volume expansion of the inner short-rod portion of the edge, but this is more than compensated by the larger decrease of accessible volume in the compressed outer long-rod portion of the interface. Overall the curved interface reduces the net volume excluded to the polymer and is thus energetically favorable when compared to the straight interface (**Fig. 2f**). Analogously, curving the interface towards the long rod rafts *will increase* the net membrane volume, since the expansion of the long rod raft is larger than the complementary compression of the short rods (**Fig. 2b**). We argue that the asymmetric expansion/compression favor interface curvature *towards* the shorter rod domains and *away* from the longer rod domains, and drastically increases the susceptibility of so-curved interfaces to edge twist.

**Edge Tilt Dependence with Raft Size**: To test these ideas we used colloidal membranes, which are one rod-length thick liquid-like monolayers of aligned rods assembled by the depletion interaction (*19-21*). Previous work showed that 880-nm length left-handed *fd*-Y21M rod-like viruses formed finite-sized colloidal rafts when dissolved in a background membrane composed of longer 1200-nm M13KO7 rods of the opposite chirality (*25, 31, 32*). When constituent rods are parallel to the membrane normal that lies in the image plane, the retardance viewed with LC-PolScope is near zero and the image appears locally dark. A local tilt of rods away from the membrane normal introduces local birefringence, which leads to increased signal in LC-PolScope images (*33, 34*). Imaging short-rod *fd*-Y21M rafts with LC-PolScope revealed a bright ring around the raft edges, which is indicative of the local tilt (**Fig. 3a, c, Movie S2, S3**). We explored the complementary regime, in which a low volume fraction of M13K07 long rods was dissolved in a membrane of short *fd*-Y21M rods. The long rods also formed finite-sized clusters, but these were more heterogeneous (**Fig. 3b, Movie S2**). Observing such clusters with LC-PolScope revealed the absence of any measurable twist of similarly sized rafts (**Fig. 3d, e**). Theories which consider only the effect of chiral asymmetry between the raft and the background domain predict that equal-size rafts of both short- and long-rod would twist by the same amount. In contrast, the near complete suppression of the edge twist in the long-rod rafts in the presence of chiral asymmetry implies that the interface structure is sensitive to the length difference between the inner and outer phase, consistent with previously outlined geometrical arguments.

To quantify these effects we extended the existing smectic-layer model of chiral colloidal rafts (*21, 23*), by incorporating the thermodynamic costs due to the hyperbolic edge geometry (Appendix A). In the smectic model, phase separated domains are described by Frank elastic gradients for bend, splay and twist along the midplane (*35*). We assume that all elastic deformations have the same elastic constant $K$. The chiral preferences of the inner and outer domain are parameterized by the inverse pitch parameters $q_{in}$ and $q_{out}$. Our model also includes a penalty for rod tilt away from the layer normal that derives from the osmotic cost ($\Pi$ per unit volume) of increasing areal exposure of a depletion layer of thickness $a$ at the top and bottom surface upon tilt while maintaining a constant inter-rod density at the midplane. A torque balance between the two effects gives equilibrium profiles of tilt that extend from the inter-domain edge by a penetration length $\lambda = \sqrt{K\ell/2\Pi a}$. Our model accounts for the changes in the excluded volume due to the compression/dilation of the hyperbolic domain edge. As discussed above, splay deformation requires variation of the inter-rod density $\rho$ in the layer away from the preferred value. We approximate this cost by a two-body term $A_2 \rho^2/2$, where $A_2$ is the second virial coefficient for inter-rod density. In combination with the rod translational entropy this balances the osmotic pressure exerted by the enveloping depletant.

Our central result is summarized by considering an expansion of the raft free energy for radius $R$ in terms the maximum interface tilt $\theta_0$:

$$F(\theta_0, R) = 2\pi R\, \gamma + 2\pi R\, \theta_0\, K\Delta(q\ell) + \frac{\chi^{-1}(R)}{2}\theta_0^2 + O(\theta_0^4). \qquad (1)$$

Here, $\gamma = \Pi a |\ell_{in} - \ell_{out}|$ is the tilt-independent line tension due to depletion at the "side wall" between the long and short domains. The next term represents the linear free energy gain due to twist in opposing sides of domain edge, which is proportional to the difference between the preferred inverse pitch of raft and membrane domains for the case of length-symmetric rods (*21,* 23), but more generally, depends on the difference of the product of rod length and chirality $\Delta(q\ell) = q_{out}\ell_{out} - q_{in}\ell_{in}$. The second order coefficient, $\chi^{-1}(R) = \chi_{sm}^{-1}(R) + \chi_{edge}^{-1}(R)$, describes the tilt stiffness, or *inverse tilt susceptibility,* and has two contributions. The previously derived smectic contribution $\chi_{sm}^{-1}(R)$ arises from the balance of elastic costs of tilt gradients and the penalty for tilt (*23*). For small rafts, the linear gradient of tilt within the raft generates a constant stiffness $\chi_{sm}^{-1}(R \ll \lambda) \approx K\ell$, while for large rafts, the localized tilt at the edge gives rise to a tilt stiffness $\chi_{sm}^{-1}(R \gg \lambda) \approx K\ell R/\lambda$. The second contribution to the tilt stiffness, $\chi_{edge}^{-1}(R)$, arises from the hyperbolic geometry of the raft/membrane edge:

$$\chi_{edge}^{-1}(R) = \alpha \Pi \, \partial_{\theta_0}^2 V_{edge} + a \Pi \, \partial_{\theta_0}^2 A_{edge}. \tag{2}$$

Here, $V_{edge}$ and $A_{edge}$ are contributions from the volume and surface area change of the membrane due to the hyperbolic raft edge shape, and $a$ is the size of the depletant (Appendix B, eqs. B12 and B13). The coefficient $\alpha$ derives from the net work of osmotic pressure in the change of edge volume and counterbalancing change in 2-body rod-rod interactions due to the splay induced compression/dilation. We estimate a value of $\alpha = \rho \frac{k_B T}{\Pi \ell} < 1$, which can be expected based from van't Hoff law prediction for experimentally relevant osmotic pressures (*36*). Expansion to the lowest order in the tilt angle yields: $V_{edge} \simeq \pi(\ell_{in}^3 - \ell_{out}^3)\theta_0^2/12$ and $A_{edge} \simeq \pi|\ell_{in}^3 - \ell_{out}^3|\theta_0^2/(12\,R) + \pi(\ell_{in}^2 - \ell_{out}^2)\theta_0^2/2 - \pi|\ell_{in} - \ell_{out}|\,R\,\theta_0^2$. The volume change results from the difference between the compression/expansion of the inner and outer membrane edge, while the area change results from the corresponding change in the hyperbolic edge surface area. For short-rod rafts $V_{edge}(\ell_{in} < \ell_{out}) < 0$, indicating that the osmotic pressure of the enveloping polymer *increases* the tilt susceptibility, whereas for long-rod rafts ($\ell_{in} > \ell_{out}$) the net dilation of the membrane volume stiffens the resistance to tilt. Note that the length-asymmetry of the tilt susceptibility is further enhanced by a factor proportional to $a(\ell_{in}^2 - \ell_{out}^2)$ deriving from the change of $A_{edge}(\ell_{in} < \ell_{out})$. Because $\chi(R)|_{\ell_{in}<\ell_{out}} > \chi(R)|_{\ell_{in}>\ell_{out}}$, the equilibrium tilt at a given raft size, $\theta_*(R) = 2\pi R\,K\Delta(q\ell)\,\chi(R)$, increases with length asymmetry of small-rod rafts and is suppressed for the long rod rafts. The latter prediction is consistent with the observed suppression of twist of long-rod rafts (**Fig. 3**).

Another central prediction of our model is the non-monotonic dependence of the maximum edge twist $\theta(R)$ of short-rod rafts as a function of raft radius R. With increasing size, the maximum angle $\theta_0(R)$ first increases, reaching a maximum for raft sizes that are of the order of the twist penetration length. Increasing the raft size further untwists the interface (**Fig. 4e**). This behavior derives from the generic evolution of twist stiffness $\chi^{-1}(R)$ with raft size. In the $R \to 0$ limit the hyperbolic edge geometry stiffens rafts with $A_{edge} \to \pi|\ell_{in}^3 - \ell_{out}^3|\theta_0^2/(12\,R)$ (**Fig. 4f,** left); at intermediate sizes the net compression of the $V_{edge}$ leads to a softened response to twist (**Fig. 4f,** middle); while for large rafts the hyperbolic geometry of twisted edges is flattened and the linear growth of $\chi_{sm}^{-1}(R \gg \lambda) \approx K\ell R/\lambda$ dominates, restiffening the raft to twist (**Fig. 4f,** right). The model predicts that the rod-length difference strongly influences the non-monotonic dependence of twist on raft size, with a peak in maximum twist occurring when the length difference is comparable to the experimental values **(Fig. 4e).** In comparison, the predicted twist for rafts as radius $r \to \infty$ is independent of $\Delta\ell$.

To test this prediction, we measured the maximum twist for different sized rafts using LC-PolScope. With increasing radius the retardance increased, reaching a maximum for ~1.5 μm diameter rafts (**Fig. 4a, b**). Beyond this size maximum retardance decreased, eventually dropping to near background levels (**Fig. 4c**). To clarify the contribution of chiral effects, we repeated measurements for the background membranes with different twisting power by mixing different ratios of left-handed (M13KO7) and right-handed (M13KO7-Y21M) (*26, 37*). Using this method, the background membranes was tuned to be either strongly left-handed ($\phi_{BG} = 1.0$) or effectively achiral ($\phi_{BG} = 0$) (**Fig. 4d**). All three cases exhibited non-monotonic behaviors, but the peak in the twist for the fully chiral background was significantly less pronounced.

**Long-Rod Raft Shape Instability**: It is illuminating to recast the thermodynamic coupling between the double twist and edge curvature in terms of an effective line energy for the edge length $s$ and its geodesic curvature $\kappa_g$. In the limit of large and slowly varying edge curvature this takes the following form:

$$E_{edge} \cong \oint ds \left[\gamma' + \frac{B}{2}(\kappa_g - \kappa_0)^2\right], \tag{3}$$

where $\gamma' \cong \gamma - \delta\gamma^*$ is the line tension renormalized by chiral edge twist, $\delta\gamma^* \approx K\ell\lambda^2(\Delta q)^2$ (Appendix B). The hyperbolic edge geometry generates an effective edge bending elasticity for length asymmetric membranes, with bending stiffness $B \cong K\ell(\Delta q)^2 |\ell_{in}^3 - \ell_{out}^3|$, and a preferred geodesic curvature $\kappa_0 = \text{sgn}(\ell_{out} - \ell_{in})\left(\frac{\alpha}{2a} + \frac{3}{|\ell_{in}-\ell_{out}|}\right)$, reflecting the preference for twisted edges to bend *towards* the short-rod domain, as is observed in short rod rafts. Considering the opposite scenario of long-rod rafts, the preferred edge curvature, $\kappa_0$, resists the generic tendency to maintain a circular shape, favored by positive $\gamma'$. Furthermore, it suggests that circular long-rod rafts will become unstable to non-convex domain shapes.

Long-rod rafts exhibited such shape instabilities. In native samples, most long-rod rafts had an area of ~1 µm² and were disk-shaped. Intriguingly, the occasionally observed larger rafts always assumed anisotropic shapes. To quantify this instability, we systematically increased the size of long-rod rafts. Weak repulsive interactions allowed us to merge long-rod rafts with optical tweezers, thus systematically increasing their size (**Fig. 5b**). After reaching an area of ~1.5 µm², such rafts assumed elongated shapes (**Fig. 5a**). Above a ~4 µm² critical area, they transformed into nonconvex horseshoe-like shapes, reminiscent of stomatocyte vesicle shapes (Movie S4) (*27*). Further size increase yielded closed and stable annularly shaped rafts, where short rods resided both inside and outside the annulus. The evolution of raft shapes was quantified by plotting the ratio of the minor to major axis of the structure. The marked decrease in the ratio indicated a transition from a circular disk to an elongated shape (**Fig. 5c**). The subsequent increase and the plateau signified the transition to horseshoe-like structures and closed annuli. LC-PolScope revealed differences in the structures of the inner and outer interfaces of the annular rafts, which correspond to the same long and short-rod interface with the opposite curvature. Rods twisted along the inner edge, and this twist was comparable in magnitude to that of a similarly-sized short-rod raft in a long-rod background (**Fig. 6a, b**). In contrast, the retardance at the annulus's outer edge was only slightly above the background level, indicating the absence of measurable twist at the outer long-rod to short-rod interface, much like the retardance measurements for long rod rafts (**Fig. 6e, 3e**). Such twist asymmetry yields different fluctuations of the inner and outer interface (Movie S5). Analogous to the free edge of a membrane (*38, 39*), the twist of the inner annular interface introduces an elastic penalty that suppresses small wavelength fluctuations. In comparison, the outer untwisted interface has no such penalty and exhibits enhanced fluctuations.

To describe these observations, we adapted the circular raft model to an annular geometry with variable inner/outer radii and tilt angles (Appendix C). Plotting the relative free energy of a compact tall-rod raft and an annular domain with a twisted, short-rod raft at its center, shows that increasing the total area of tall-rod rafts reduces the tilt-mediated edge-edge repulsions in the annular domain (**Fig. 5e**). For sufficiently large area, this effect, in combination with the incorporation of an inner edge with the preferred direction of geodesic curvature, stabilizes annuli over disks (**Fig. 5d**).

**Discussion and conclusions**: Using a combination of experiments and theory we have described a generic geometric coupling between the orientation and composition of phase separated fluid membranes. We demonstrated that the length asymmetry of the constituent rods gives rise to a novel mechanism of domain shape selection, that is driven by a spontaneous non-zero curvature of the interface towards the shorter rod domain. The consequence of spontaneous curvature have been extensively studied for analogous 3D vesicle, where locally-selected edge curvature generated by the asymmetry in the two bilayer leaflets drives large scale shape changes (*27, 40, 41*). Membrane embedded fluid droplets are similar to 3D vesicles in that they can assume multiple polymorphic shapes. However, there are also important differences in that for 2D droplets the spontaneous curvature is determined by the properties of the phase separated fluids, while in 3D vesicles the curvature is selected by the structural asymmetry of the lipid bilayer itself.

We focused on the regime of comparable lengths where the tilt stiffness $\chi^{-1}(R) > 0$ for all raft sizes, such that the edge twist only arises due to chiral asymmetry in the mixture (i.e. when $\Delta q \neq 0$). However, we anticipate that above a critical length difference the edge-compression mechanism will lead to negative twist stiffness ($\chi^{-1}(R) < 0$) for a range of intermediate raft sizes. In such a regime, edges would spontaneously twist even in the absence of chirality. Furthermore, for sufficiently small chiral bias, negative twist susceptibility should give rise to metastable rafts with twist that is opposite of the native one. Intriguingly, such counter twisted rafts have been observed in membranes with a low net chirality (*26*). Furthermore, initial observations indicate that raft twist in fully achiral membranes is comparable or greater than the chiral membrane (**Fig. 7**). While here we focused on the specific case of colloidal monolayers, the mechanism that generates spontaneous curvature can be recast as a generic consequence of the coupling between the *compositional* and *orientational* degrees of freedom, that should be relevant for all phase-separated fluid membranes with tilt degrees of freedom. These effects may be relevant to the broader class of fluid membrane systems, leading to a rich variety of 2D fluid domain shape equilibria.

Multiple mechanism can lead to assembly of finite-sized fluid assemblages. For example, surfactants and other amphiphilic molecules assemble into micelles and other finite-sized structures, but in these systems the final assemblage size is always comparable to the size of the molecular constituents (*42*). A closely related class of cluster forming system, exemplified by magnetic fluids, is based on microscopic units with repulsive interactions whose range in much longer than the particle size. Such systems can form droplets with complex shapes, but again the cluster size is determined by the range of the repulsive potential, which is much larger than the constituent elements (*43, 44*). An alternate self-closing mechanism that leads to finite-sized assemblages is exemplified by virus capsids, wherein a predetermined curvature emerges from the interactions of the microscopic building blocks, and the final assemblage size is determined by this curvature (*45*). A final mechanism is based on geometrical frustration, in which the finite size is determined by the fundamental incompatibility between local interactions and the global assembly constraints (*46*). Previous work has suggested that assembly of colloidal rafts, which typically contain tens of thousands of rodlike molecules, and are thus much larger than any molecular parameter, belong to the last class of geometrically frustrated assemblages. In particular, modeling suggested that their size is determined between the inherent frustration between the local preference of rods to twist which in incompatible with the global constrains of assembly of 2D sheets which expel twist (*23, 24*). However, work described here demonstrates presence of an additional previously unidentified self-closing mechanism, wherein the geometrical properties of the demixed fluid phases in length asymmetric molecules chooses a preferred edge curvature, which in turn selects and stabilizes finite-sized assemblages.

**Appendix A: Membrane Model Energy**

To describe the osmotically-condensed rod assembly, we build on the previously developed theoretical model for a liquid crystalline membrane that was used to describe chirality-driven rafts structure in this system (*23*). The model free energy is:

$$F = F_{Frank} + F_{Osmotic} + F_{Compression} + F_{Ideal}. \quad (A1)$$

These terms account for rod orientational interactions, depletant excluded volume changes due to the membrane, rod density-dependent interactions, and the 2D rod translational entropy, respectively. The penalties for gradients in rod orientation **n** are captured as in the previous work by a chiral nematic Frank elastic energy,

$$F_{Frank} = \frac{K}{2} \int dV[(\nabla \cdot \mathbf{n})^2 + (\nabla \times \mathbf{n})^2 - 2q(\mathbf{n} \cdot \nabla \times \mathbf{n})], \quad (A2)$$

where $q$ is the preferred chiral pitch and $K$ is the elastic constant. The second term is due to the translational entropy of the depletant, $F_{Osmotic} = \Pi V_{ex}$, with $V_{ex}$ the change in volume available to the depletant and $\Pi$ the osmotic pressure on the rods due to the depletant (*47*). For a spherical depletant of radius $a$, $V_{ex}$ is the volume of space within a distance of $a$ from the rods from which the depletant is excluded. It was previously shown that this entropic effect can drive phase separation and formation of colloidal rafts (*23*). Upon phase separation, this term penalizes tilt, limiting twist to the boundaries of membrane rafts with twist penetration $\lambda$ set by the balance of $F_{Osmotic}$ and $F_{Frank}$ (*34*). We additionally consider the role of rod splay away from the midplane, which derives from inter-domain edge twist and modifies $V_{ex}$ from its previous form (*23*). We also consider the energy due to expansion/compression in the rod areal density $\rho$, $F_{Compression} = \frac{A_2}{2} \int dV \rho^2$, with $A_2$ being the second virial coefficient. Critically, this density $\rho$ is determined by the area packing of rods *perpendicular to their local tangent* **n**, as this density characterizes the local distances of closest approach between points on distinct rods (*48, 49*). This term is treated separately from the Frank elastic energy, which considers rod packing at the midplane

Lastly, the ideal rod translational entropy is $F_{Ideal} = k_B T \int dA\, \rho (\log \rho - 1)$. The 2D translational entropy of rod centers contributes to the preferred rod spacing: for the untwisted case in a domain of rods with rod length $\ell$, $\left(\frac{\partial F}{\partial A}\right)_N = \Pi(\ell + 2a) - \frac{A_2}{2}\ell\rho^2 - k_B T \rho$, and mechanical equilibrium determines the rod density to be $\rho_0$ such that $\left(\frac{\partial F}{\partial A}\right)_N = 0$. From this, the second virial coefficient can be related to measurable parameters including the straight-rod areal density $\rho_0$ by rearranging this condition to obtain:

$$\frac{1}{2}A_2\rho_0^2 \approx \Pi - \frac{k_B T \rho_0}{\ell}, \quad (A3)$$

Upon twisting, we consider variation of rod density with the assumption that the areal density at the midplane remains constant as measured perpendicular to the rods, $\rho = \rho_0 \cos\theta$. Notably, this requires the lateral spreading of tilted domains described previously (*23*). However, we recognize that it is not possible to enforce constant density throughout the membrane volume due to edge splay. The small change in $F_{Ideal}$ as the fixed-rod number membrane dilates (while maintaining $\rho = \rho_0 \cos\theta$), leads to a neglible contribution for the free energy at small tilt. To quadratic order in the tilt, one finds that there is zero splay at the midplane $z = 0$ for the geometry considered below, but there is nonzero splay and a spatially varying density for $z \neq 0$ as described below. The variational model is minimized over the rod tilt profile $\theta(r)$, and the additional degrees of freedom are the boundary radii of the structures considered, either the radius of the raft $R$ or the inner and outer radii of the annulus structure $R_1, R_2$.

**Appendix B: Single Raft Harmonic Model**

In this section, the raft size-dependent tilt susceptibility $\chi^{-1}$ (Eqs. 1, 2 in the main text) is derived to predict the chirality-driven equilibrium raft tilt at the boundary, $\theta_0 \approx K\ell\Delta qR / \chi^{-1}$, accurate to quadratic order in $\theta_0$ and first order in the depletant size $a$. The raft domain is characterized by rods of length $\ell_{in}$, rods in the surrounding membrane have length $\ell_{out}$, and the domain boundary is located at $r_0 = R$. Near the domain boundary, rods tilt away from the layer normal and perpendicular to the radial direction, so their orientations are described by:

$$\mathbf{n}(r_0, \phi) = \cos\theta(r_0)\,\hat{z} + \sin\theta(r_0)\,\hat{\phi}, (B1)$$

where $\theta(r_0)$ is the rod tilt profile. Rod tilt is localized near the boundary to within a characteristic twist penetration length defined below. The "isolated" raft structure is embedded in a background membrane whose outer radius is taken to $r_0 \to \infty$. The rod central axes are then described by the family of curves $\mathbf{r}(r_0, z_0, \phi) = r_0 \hat{r} + z_0 \mathbf{n}(r_0, \phi)$. The material coordinate $r_0$ is the radial coordinate of a rod center in the midplane (Fig. 1e), and the material coordinate $z_0$ is the distance measured along the rod axis. The excluded volume is evaluated to linear order in the depletant radius $a$ using the Jacobians that map material coordinates $(r_0, z_0, \phi)$ to spatial coordinates $\mathbf{r}$:

$$V_{ex} = \int dr_0 \, dz_0 \, d\phi \, J_{\text{bulk}} + a \int dr_0 d\phi \, g_{\text{top}} + a \int dz_0 \, d\phi \, g_{\text{edge}} + O(a^2) \quad (B2)$$

where

$$J_{\text{bulk}} = r_0 \cos\theta \left(1 + \frac{z_0^2}{r_0} \tan\theta \, \partial_{r_0}\theta\right), (B3)$$

$$g_{\text{top}} = \sqrt{r_0^2 + \left(\frac{\ell}{2}\right)^2 \partial_{r_0}\theta \left(r_0 \sin 2\theta + \left(r_0^2 + \left(\frac{\ell}{2}\right)^2\right) \sin^2\theta \, \partial_{r_0}\theta\right)}, (B4)$$

$$g_{\text{edge}} = \sqrt{\frac{1}{2}(R^2 + z_0^2 + (R^2 - z_0^2) \cos 2\theta)}. (B5)$$

The first integral, $V_{in} + V_{out}$, is over the "bulk" volume of the membrane, the second $A_{in} + A_{out}$ is over the surface of exposed rod ends with $z_0 = \pm \ell_{in}/2$ for the raft and $z_0 = \pm \ell_{out}/2$ in the surrounding membrane, and the third integral $A_{edge}$ is over the exposed rod sides at the domain boundary $r_0 = R$ with $z_0$ varying from the the top of the raft edge to the top of the surrounding membrane edge, i.e. from $z_0 = \ell_{in}/2$ to $\ell_{out}/2$ and $z_0 = -\ell_{in}/2$ to $-\ell_{out}/2$ (Fig. 8).

The density at midplane is fixed to be $\rho = \rho_0 \cos\theta$. Away from the midplane, $\frac{\rho}{\rho_0} = \frac{J_{\text{bulk}}(r_0, 0, \phi)}{J_{\text{bulk}}(r_0, z_0, \phi)} = \left(1 + \frac{z_0^2}{r_0} \tan\theta \, \partial_{r_0}\theta\right)^{-1}$. This distribution is consistent with the conservation of rod flux (i.e. $\nabla \cdot (\rho \mathbf{n}) = 0$) in the membrane interior. The splay distortion, given by:

$$\nabla \cdot \mathbf{n} = \frac{2 z_0 \, \partial_{r_0}\theta}{z_0^2 \, \partial_{r_0}\theta + r_0 \cot\theta}, (B6)$$

is non-zero near the inter-domain edge where $\partial_{r_0}\theta \neq 0$. It can be verified that the exact splay distribution agrees with the simplified limit presented in the main text, $\nabla \cdot \mathbf{n} \approx z \, \partial_r(\tan^2\theta)/r$, for small tilt. Conservation of the rod number in the raft, in combination with the fixed inter-rod density at the midplane, leads to the shift of the domain boundary from $R$ to $r_0 = R'$. This constraint requires that $\int_0^R dr_0 \, r_0 = \int_0^{R'} dr_0 \, r_0 \cos\theta$.

We expand this theory to second order in the edge tilt $\theta(r_0)$. To this order the tilt-dependent excluded volume $V_{ex}$ has terms associated with the raft boundary that only depend on the tilt at the boundary $\theta_0 = \theta(R')$, combined with a "smectic"-like coupling (favoring normal rod alignment) due to the requisite expansion of rafts upon tilt, displacing the interdomain boundary, and also the outer boundary at $r_0 \to \infty$,

$$\Delta F_{sm,in} = \Pi a \left( 2\pi R'^2 + 2\pi R' |\ell_{in} - \ell_{out}| - 2\pi R^2 - 2\pi R |\ell_{in} - \ell_{out}| \right)$$
$$\simeq \Pi a 4\pi \left( \int_0^R dr_0 \, r_0 \frac{\theta^2}{2} \left( 1 + \frac{|\ell_{in} - \ell_{out}|}{R} \right) \right),$$

$$\Delta F_{sm} \simeq \Pi a 4\pi \left( \int_0^R dr_0 \, r_0 \frac{\theta^2}{2} \left( 1 + \frac{|\ell_{in} - \ell_{out}|}{R} \right) + \int_R^\infty dr_0 \, r_0 \frac{\theta^2}{2} \right). \quad (B7)$$

where $\Delta F_{sm,in}$ describes the free energy change associated with the raft domain and its vertical edge. Also, to second order in the rod tilt $\theta(r_0)$, the Frank elastic energy cost is due to the varying twist in the midplane,

$$F_{Frank} \simeq \pi K_{in} \ell_{in} \left( -2 q_{in} R \theta_0 + \theta_0^2 + \int_0^R dr_0 \, \frac{\theta^2}{r_0} + r_0 (\partial_{r_0} \theta)^2 \right)$$
$$+ \pi K_{out} \ell_{out} \left( 2 q_{out} R \theta_0 - \theta_0^2 + \int_R^\infty dr_0 \, \frac{\theta^2}{r_0} + r_0 (\partial_{r_0} \theta)^2 \right). \quad (B8)$$

Together, $F_{Frank}$ and $\Delta F_{sm}$ are the only terms that are not associated with the inter-domain edge in the harmonic model, and these non-edge domain energies together restrict twist to the boundary with a characteristic length $\lambda = \sqrt{K\ell/2\Pi a}$ as derived previously (*23*). More precisely, the variation of $F_{Frank} + \Delta F_{sm}$ with respect to $\theta(r_0)$ and fixed boundaries $\theta(R') = \theta_0$ and $\theta(0) = \lim_{r_0 \to \infty} \theta(r_0) = 0$ gives:

$$\theta(r_0) = \begin{cases} \theta_0 \dfrac{I_1(r_0/\lambda_{in}(R))}{I_1(R/\lambda_{in}(R))}, & 0 < r_0 < R \\ \theta_0 \dfrac{K_1(r_0/\lambda_{out})}{K_1(R/\lambda_{out})}, & R < r_0 \end{cases} \quad (B9)$$

where $I_1$ is the first-order modified Bessel function of the first kind and $K_1$ is the first-order modified Bessel function of the second kind. The penetration depths of the tilt into the inner and outer domain are given by:

$$\lambda_{in}^{-1}(R) = \sqrt{\frac{2\Pi a \left(1 + \frac{|\ell_{in} - \ell_{out}|}{R}\right)}{K_{in}\ell_{in}}} \quad (B10)$$

$$\lambda_{out}^{-1} = \sqrt{\frac{2\Pi a}{K_{out}\ell_{out}}}. \quad (B11)$$

With this twist profile we expand $F = F(R, \theta_0)$ to quadratic order in $\theta_0$ and first order in depletant radius $a$.

Deriving from the hyperbolic membrane geometry, the change in membrane volume is

$$\Delta V_{edge} = \Delta \int dr_0 \, dz_0 \, d\phi \, J_{\text{bulk}} = \frac{\pi(\ell_{in}^3 - \ell_{out}^3)\theta_0^2}{12} + O(\theta_0^4). \, (B12)$$

Osmotic pressure favors decreasing the bulk membrane volume $\Pi \Delta V_{edge}$. Counteracting this effect is the change of the two-body rod interaction energy due to tilt induced local compression/expansion away from the preferred rod density:

$$\Delta F_{Compression} \simeq -\frac{A_2}{2}\rho_0^2 \frac{\pi(\ell_{in}^3 - \ell_{out}^3)\theta_0^2}{12}. \, (B13)$$

Together, $\Pi \Delta V_{edge} + \Delta F_{Compression} \simeq \alpha \Pi \Delta V_{edge}$ with $\alpha = 1 - \frac{A_2 \rho_0^2}{2\Pi}$. From the mechanical equilibrium of straight-rod domains considered in the previous section, for $\ell \gg a$, $\Pi \ell - \frac{A_2}{2}\ell\rho_0^2 - k_B T \rho_0 \approx 0$ so that $\alpha = \frac{k_B T \rho_0}{\Pi \ell}$, which is estimated numerically below.

The remaining tilt-dependent excluded volume terms are due to the change in the surface area of the membrane $a\Pi A_{edge}$, with the edge area equal to:

$$\Delta A_{edge} = \Delta \int dz_0 \, d\phi \, g_{\text{edge}} = \pi \frac{|\ell_{in}^3 - \ell_{out}^3|}{12 \, R}\theta_0^2 + \pi \frac{\ell_{in}^2 - \ell_{out}^2}{2}\theta_0^2 - \pi|\ell_{in} - \ell_{out}| R \, \theta_0^2 + O(\theta_0^4). \, (B14)$$

The first term is due to the increasingly curved domain edge with increasing tilt, the second is due to the radial displacement of the edge, and the final one is due to the vertical contraction of the edge.

The tilt-dependent terms are added to the untilted line tension $\gamma = \Pi a |\ell_{in} - \ell_{out}|$ to capture both the size- and tilt-dependence of the raft membrane energy:

$$F(\theta_0, R) = 2\pi R \gamma + 2\pi R \theta_0 (\ell_{out} K_{out} q_{out} - \ell_{in} K_{in} q_{in}) + \frac{\chi^{-1}(R)}{2} \theta_0^2, \quad (B15)$$

where the inverse tilt susceptibility of the raft is:

$$\chi^{-1}(R) = \alpha \Pi \, \partial_{\theta_0}^2 V_{edge} + a \Pi \, \partial_{\theta_0}^2 A_{edge}$$

$$+ \partial_{\theta_0}^2 \left( \int_0^R dr_0 \, r_0 \left( \pi K_{in} \ell_{in} \left( \frac{\theta}{r_0} + \partial_{r_0} \theta \right)^2 + \Pi a 4\pi \frac{\theta^2}{2} \left( 1 + \frac{|\ell_{in} - \ell_{out}|}{R} \right) \right) \right.$$

$$\left. + \int_R^\infty dr_0 \, r_0 \left( \pi K_{out} \ell_{out} \left( \frac{\theta}{r_0} + \partial_{r_0} \theta \right)^2 + \Pi a 4\pi \frac{\theta^2}{2} \right) \right). \quad (B16)$$

Here, the final two terms, defined $\chi_{Sm}^{-1}(R)$ in the main text, are identical to the previously derived form (23), with the exception of a small numerical correction of $\frac{|\ell_{in} - \ell_{out}|}{R}$ in the form of a raft-radius dependence of the inner-domain penetration length, $\lambda_{in}(R)$.

Given these expressions, a prediction for the optimal edge tilt is:

$$\theta_0^*(R) = \frac{2\pi (\ell_{out} K_{out} q_{out} - \ell_{in} K_{in} q_{in}) R}{\chi^{-1}(R)} \quad (B17)$$

where we have assumed: i) the limit of small tilt where higher-order $\theta_0$ corrections are neglible and ii) that $\chi^{-1}(R) > 0$ so that tilt stiffness remains finite at all raft sizes.

To develop an effective theory for the edge shape, we consider the limit of small curvature with negligible costs associated with the variable edge tilt $\theta(s)$, where $s$ is the arc coordinate of the edge. Identifying $ds = R \, d\phi$ and geodesic curvature $\kappa_g = R^{-1}$, we rewrite the harmonic theory as a function of edge shape and tilt:

$$E_{edge}[\theta(s)] \cong \oint ds \left[ \gamma - \gamma_1 \theta + \frac{\gamma_2}{2} \theta^2 + \frac{b\theta^2}{2} (\kappa_g - \kappa_0)^2 \right], (B18.A)$$

where the coefficients were derived from SI eqs. (18, 19) in the large $R$ limit

$$\gamma_1 = (\ell_{in} K_{in} q_{in} - \ell_{out} K_{out} q_{out}), \quad (B18.B)$$

$$\gamma_2 = a\Pi (2\lambda_{in} + 2\lambda_{out} - |\ell_{in} - \ell_{out}|) - b\kappa_0^2, \quad (B18.C)$$

$$b = \frac{a\Pi}{12}|\ell_{in}^3 - \ell_{out}^3|, \text{(B18.D)}$$

$$\kappa_0 = -\left(\frac{\alpha}{2a} + \frac{3}{|\ell_{in} - \ell_{out}|}\right)\text{sgn}(\ell_{in} - \ell_{out}). \text{(B19)}$$

Minimizing over edge tilt and considering a limit of small bending energy (strictly justified in the limits of small $\alpha$ or vanishing length asymmetry), we find the form of the main text eq. 3, with a renormalized line tension:

$$\gamma' \cong \gamma - \frac{\gamma_1^2}{2\gamma_2} \text{(B20)}$$

and edge bending modulus

$$B \cong \frac{\gamma_1^2 b}{2\gamma_2^2}. \text{(B21)}$$

**Appendix C: Annulus Harmonic Model**

The annulus has circular boundaries at $r_0 = R_1$ and $r_0 = R_2$, so that the annular domain of long rods has area $A_{annulus} = \pi R_2^2 - \pi R_1^2$. Tilting of rods is permitted at both boundaries, $\theta(R_1) = \theta_1$ and $\theta(R_2) = \theta_2$, and again $\theta(0) = \lim_{r_0 \to \infty} \theta(r_0) = 0$. The annular domain has rods of length $\ell_{long}$ and the membrane regions inside and outside have rods of length $\ell_{short}$. To determine the optimal inner radius, the free energy of the annulus is taken with respect to a long-rod raft of negligible tilt and area $A_{annulus}$:

$$\Delta F = 2\pi \left(R_1 + R_2 - \sqrt{R_2^2 - R_1^2}\right)\gamma$$
$$+ 2\pi\left(\ell_{long}K_{long}q_{long} - \ell_{short}K_{short}q_{short}\right)(R_1\theta_1 - R_2\theta_2) + \frac{\chi_{11}^{-1}}{2}\theta_1^2$$
$$+ \chi_{12}^{-1}\theta_2\theta_1 + \frac{\chi_{22}^{-1}}{2}\theta_2^2 \text{ (C1)}$$

where:

$$\chi_{11}^{-1} = a\Pi\pi \frac{(\ell_{short}^3 - \ell_{long}^3)}{6} + a\Pi\pi \left(\frac{|\ell_{long}^3 - \ell_{short}^3|}{6 R_1} + (\ell_{short}^2 - \ell_{long}^2)\right.$$
$$\left. - 2|\ell_{long} - \ell_{short}| R_1\right)$$
$$+ \partial_{\theta_1}^2 \left(\int_0^{R_1} dr_0\, r_0 \left(\pi K_{short}\ell_{short}\left(\frac{\theta}{r_0} + \partial_{r_0}\theta\right)^2\right.\right.$$
$$\left.\left. + \Pi a 4\pi \frac{\theta^2}{2}\left(1 + \frac{|\ell_{long} - \ell_{short}|}{R_1} + \frac{|\ell_{long} - \ell_{short}|}{R_2}\right)\right)\right.$$
$$\left. + \int_{R_1}^{R_2} dr_0\, r_0 \left(\pi K_{long}\ell_{long}\left(\frac{\theta}{r_0} + \partial_{r_0}\theta\right)^2 + \Pi a 4\pi \frac{\theta^2}{2}\left(1 + \frac{|\ell_{long} - \ell_{short}|}{R_2}\right)\right)\right.$$
$$\left. + \int_{R_2}^{\infty} dr_0\, r_0 \left(\pi K_{short}\ell_{short}\left(\frac{\theta}{r_0} + \partial_{r_0}\theta\right)^2 + \Pi a 4\pi \frac{\theta^2}{2}\right)\right), \text{(C2)}$$

$$\chi_{12}^{-1} = \partial_{\theta_1}\partial_{\theta_2}\left(\int_0^{R_1} dr_0\, r_0\left(\pi K_{short}\ell_{short}\left(\frac{\theta}{r_0} + \partial_{r_0}\theta\right)^2\right.\right.$$
$$\left.\left. + \Pi a 4\pi \frac{\theta^2}{2}\left(1 + \frac{|\ell_{long} - \ell_{short}|}{R_1} + \frac{|\ell_{long} - \ell_{short}|}{R_2}\right)\right)\right.$$
$$\left. + \int_{R_1}^{R_2} dr_0\, r_0 \left(\pi K_{long}\ell_{long}\left(\frac{\theta}{r_0} + \partial_{r_0}\theta\right)^2 + \Pi a 4\pi \frac{\theta^2}{2}\left(1 + \frac{|\ell_{long} - \ell_{short}|}{R_2}\right)\right)\right.$$
$$\left. + \int_{R_2}^{\infty} dr_0\, r_0 \left(\pi K_{short}\ell_{short}\left(\frac{\theta}{r_0} + \partial_{r_0}\theta\right)^2 + \Pi a 4\pi \frac{\theta^2}{2}\right)\right), \text{(C3)}$$

$$\chi_{22}^{-1} = \alpha \Pi \pi \frac{(\ell_{long}^3 - \ell_{short}^3)}{6}$$

$$+ a\Pi\pi \left( \frac{|\ell_{long}^3 - \ell_{short}^3|}{6 R_2} + (\ell_{long}^2 - \ell_{short}^2) - 2|\ell_{long} - \ell_{short}| R_2 \right)$$

$$+ \partial_{\theta_2}^2 \left( \int_0^{R_1} dr_0\, r_0 \left( \pi K_{short} \ell_{short} \left(\frac{\theta}{r_0} + \partial_{r_0}\theta\right)^2 \right. \right.$$

$$+ \Pi a 4\pi \frac{\theta^2}{2}\left(1 + \frac{|\ell_{long} - \ell_{short}|}{R_1} + \frac{|\ell_{long} - \ell_{short}|}{R_2}\right)\right)$$

$$+ \int_{R_1}^{R_2} dr_0\, r_0 \left( \pi K_{long} \ell_{long} \left(\frac{\theta}{r_0} + \partial_{r_0}\theta\right)^2 + \Pi a 4\pi \frac{\theta^2}{2}\left(1 + \frac{|\ell_{long} - \ell_{short}|}{R_2}\right)\right)$$

$$+ \int_{R_2}^{\infty} dr_0\, r_0 \left( \pi K_{short} \ell_{short} \left(\frac{\theta}{r_0} + \partial_{r_0}\theta\right)^2 + \Pi a 4\pi \frac{\theta^2}{2} \right)\right). \quad (C4)$$

Analogous to the analysis for a single raft, we find the optimal edge tilts,

$$\theta_1^*(R) = -2\pi(\ell_{long} K_{long} q_{long} - \ell_{short} K_{short} q_{short})(\chi_{11} R_1 - \chi_{12} R_2) \quad (C5)$$

and

$$\theta_2^*(R) = -2\pi(\ell_{long} K_{long} q_{long} - \ell_{short} K_{short} q_{short})(\chi_{21} R_1 - \chi_{22} R_2) \quad (C6)$$

where $\chi_{IJ}$ is the inverse of stiffness matrix $\chi_{IJ}^{-1}$. From this, we have the free energy of optimally twisted annuli

$$\Delta F_*(R_1, R_2) = 2\pi \left( R_1 + R_2 - \sqrt{R_2^2 - R_1^2} \right)\gamma$$

$$-2\pi^2 \left(\ell_{long} K_{long} q_{long} - \ell_{short} K_{short} q_{short}\right)^2 (\chi_{11} R_1^2 - 2\chi_{12} R_1 R_2 + \chi_{22} R_2^2) \quad (C7)$$

To construct the raft vs. annuli phase diagram (Fig. 4.d,e), the total area of tall rod domains $A_{annulus}$ is held fixed, such that $R_2$ becomes a function of inner domain radius $R_1$ via the constraint $R_2 = \sqrt{(A_{annulus} - \pi R_1^2)/\pi}$. For sufficiently large values of length asymmetry and $A_{annulus}$, a second local minimum appears at non-zero inner raft radius $R_1^*$, which corresponds to the (meta)stable annulus state. The phase boundary between compact tall-rod rafts and annuli is determined by comparing the free energy at $R_1^*$ to the free energy in the absence of internal rafts, $R_1 \to 0$.

**Appendix D: Nonlinear Model**

To assess the relative importance of non-linear corrections to the harmonic tilt theory, we numerically compute the exact volume and compressional free energy costs in the model for arbitrary edge tilt. The compression free $F_{Compression} = \frac{A_2}{2} \int dV \rho^2$ can be computed by substituting the density expression from above, $\rho = \rho_0 \cos \theta \left(1 + \frac{z_0^2}{r_0} \tan \theta \ \partial_{r_0} \theta \right)^{-1}$. The excluded volume is found by numerically constructing a radial profile of the membrane structure (**Fig. 8**). The form of the raft twist profile is as above from the harmonic model for a given fixed value of $\theta_0$, and the domain radii are shifted according to the expression above for rod number conservation, $\int_0^R dr_0 \, r_0 = \int_0^{R'} dr_0 \, r_0 \cos \theta$, the nonlinear volume is evaluated to an outer boundary at $R + 10\lambda$, which additionally shifts upon tilting to preserve outer rod number. To evaluate the volume, the envelope of the membrane rods is described by $\mathbf{r}(r_0, z_0, \phi)$ from above, from which a two-dimensional family of curves in the cross section is $\langle r, z \rangle = \langle \sqrt{r_0^2 + z_0^2 \sin^2 \theta}, z_0 \cos \theta \rangle$. The raft profile corresponds to the curves obtained by fixing $z_0 = \ell_{in}/2$ for $0 < r_0 < R'$ and $z_0 = \ell_{out}/2$ for $R' < r_0$.

From the curves describing the outer envelope of the rods, the final profile is found by shifting the envelope by the depletant size $a$, numerically finding intersections between the shifted short rod envelope and the shifted rod side envelope, and connecting the side envelope to the long rod envelope using a circular arc that is tangent at the respective endpoints (**Fig. 8**). Numerical minimization of the full free energy expression with respect to $\theta_0$ yields the equilibrium edge tilt, which is compared to the experimental measurement of the raft retardance in Fig. 3.

**Appendix E: Model Parameters**

For evaluating model predictions, we consider parameter values with a focus on the role of length-asymmetry of the model in shaping the asymmetric response to twist in the short- and long-rod rafts. Hence, we consider the simplified case of equal penetration lengths, $\sqrt{\frac{K_{long}\ell_{long}}{2\Pi a}} = \sqrt{\frac{K_{short}\ell_{short}}{2\Pi a}} = \lambda = 1$ µm, a value consistent with experimental measurements of edge twist (*37*). The effect of $\lambda_{long} < \lambda_{short}$ (anticipated due to length dependence) could somewhat decrease the tilt suppression relative to this simplifying limit, but cannot explain the non-monotonic twist behavior of the short raft with increasing radius, nor the annulus shape transition. Following the previous modeling efforts, we take the right-handed short-rod nematic pitch to be $q_{short} = 0.11$ µ$m^{-1}$, and the left-handed long-rod pitch to be $q_{long} = -0.385$ µ$m^{-1}$ (*23*). We also take model rod length values to be $\ell_{short} = 0.860$ µm and $\ell_{long} = 1.120$ µm (*23*). The effect of varying rod length is considered by fixing $\ell_{short} = 0.860$ µm and varying $\ell_{long} = \ell_{short} + \Delta\ell$. For the depletant radius we assume a value of $a = 5$ nm, $\alpha = 0.3$ which is roughly consistent with order of magnitude estimates deriving from a van't Hoff value of $\Pi = k_B T\, n_{dep} \approx 2\ 10^{-16}$ J/µ$m^3$ and $\rho_0 \approx 10^4$ 1/µ$m^2$ for experiment conditions.

From the model prediction for raft boundary tilt $\theta_0$, the peak value of optical retardance is predicted to compare against experiments. The optical retardance is related to the rod tilt in a flat membrane by $D = \Delta n\, \ell \sin^2 \theta$ where we take the birefringence to be $\Delta n_{long} = \Delta n_{short} = 0.011$ as in (*34*). The peak retardance is taken from the model retardance profile smoothed by convolution with a Gaussian of width 0.13 µ$m$ to model the limited optical resolution,

$$D_{peak} = \int_0^\infty dr_0\, r_0\, \Delta n\, \ell \sin^2 \theta\, \exp\left(\frac{|r_0 - R|}{0.13}\right)^2 . (E1)$$

**Appendix F: Colloidal Membrane Sample Preparation**

We used filamentous phages M13KO7 and *fd*. Both have ~7 nm diameters and a 2.8 µm persistence length. M13KO7 is 1200 nm-long and *fd* is 880 nm long (*25*). We also used the above phages with a single amino acid mutation in the major coat protein denoted as Y21M. This mutation yielded phages with opposite right-handed twist and an increased persistence length of 9.9 µm. All viruses were grown in liquid bacteria cultures and purified through multiple rounds of centrifugation, following previously established protocols (*50*). To eliminate the presence of longer multimers the virus was prepared at isotropic-nematic coexistence and only the isotropic phase was used for assembly of colloidal membranes as described elsewhere (*19*). Virus monodispersity was verified through gel electrophoresis.

All samples were prepared in 20 mM tris-HCl, 100 mM NaCl pH 8.0 (buffer). In the presence of depleting polymer (550 kDa Dextran), the viruses formed colloidal membranes which were observed in sealed flow chambers which were prepared with ~100 μm thick spacers. The chamber's top and bottom surfaces were coated with an acrylamide brush to prevent membrane adhesion (*51*). We labeled M13KO7 with fluorescent dye (DyLight 550 NHS ester for long rod raft experiments and DyLight 488 NHS ester dye for short rod raft experiments) with about 200 dye molecules per virus (*52*). Short-rod rafts were assembled in a mixture of 18% *fd*-Y21M and 82% M13KO7 (*25*). Long-rod rafts were studied in mixture of 13% M13KO7 and 87% *fd*-Y21M at ~33 mg/mL Dextran.

Previous experiments found that colloidal rafts were highly monodisperse (*25*). Increasing the dextran concentration from 40.0 mg/ml to 40.5 mg/ml yielded polydisperse rafts which were stable over weeks without disrupting the edge twist. For the measurements of retardance as a function of radius, the short rod raft samples were assembled at 40.5 mg/ml dextran concentration.

For short rod rafts, the chirality of the background membrane was adjusted by mixing the left-handed M13KO7 with the right-handed M13KO7-Y21M (*37*). A mixture of 37% M13KO7 and 63% M13KO7-Y21M shows no evidence of spontaneous twist (*26*). We define $\phi_{\text{BG}} = \frac{n_{M13KO7} - 0.37}{0.63}$ where $n_{M13KO7} = \frac{N_{M13KO7}}{N_{M13KO7} + N_{M13KO7\,Y21M}}$. $\phi_{BG} = 0$ corresponds to an achiral membrane and $\phi_{BG} = 1$ corresponds to maximally left-handed. Similarly, we define $\phi_{\text{RFT}} = \frac{0.26 - n_{fdwt}}{0.74}$ where $n_{fdwt} = \frac{N_{fd\,wt}}{N_{fd\,wt} + N_{fd\,Y21M}}$ so that $\phi_{Rft} = -1.0$ corresponds to maximally right-handed samples (*37*). To examine the influence of membrane chirality we studied background membranes with $\phi_{\text{BG}} = 0.0, 0.4$, and $1.0$ while fixing $\phi_{RFT} = -1.0$. We also studied the case in which both the membrane chirality and the raft chirality were minimized, $\phi_{\text{BG}} = 0.0$ and $\phi_{\text{RFT}} = 0.0$.

**Appendix G: Optical Microscopy Methods**:

Colloidal membranes were studied using a combination of differential interference contrast (DIC), phase contrast, and fluorescence microscopy. We used an inverted microscope (Nikon Eclipse TE2000-U) equipped with with oil immersion objective (Plan Fluor, 1.3 NA 100x). For fluorescence microscopy, we used Semrock FITC and TRITC filter cubes. For fluorescence excitation we used the Lumencor Sola light engine. Images were recorded with a sCMOS camera (Neo, Andor) that was controlled by MicroManager software. The local membrane tilt was measured using LC-PolScope (*33*). LC-PolScope yields images in which the pixel intensity is proportional to local sample retardance. For a flat membrane in the image plane, it is possible to translate retardance into local rod tilt, as described elsewhere (*34, 53*).

To quantify the shape dependence of long-rod rafts on their sizes, we systematically increased raft size by merging together multiple smaller rafts with optical tweezers. In comparison to short-rods rafts which are repelled away from a trap (*25, 26*), long-rod rafts could be directly trapped with a focused beam. We used a time-shared optical tweezer based on acousto-optic deflector (AOD). The traps were controlled with custom LabView software, while the imaging and microscope were controlled with Micro-Manager software. Upon merging two droplets, the resulting shape fluctuations were recorded over several minutes. We repeated the merging process several times to capture the entire transition from a roughly circular droplet to an annulus. We repeated the transition multiple times across different samples, and found the shape dependence on long rod domain size was consistent across all runs and independent of the microscopy observation techniques.

**Appendix H: Data Analysis**

For all quantitative LC-PolScope measurements, we tracked rafts in MATLAB using standard particle tracking techniques (*54*). The radial retardance profile of each short-rod raft was first radially and then temporally averaged to minimize noise. Subsequently, the average background level retardance was subtracted. We measured the retardance of more than 30 rafts of various sizes. The averaged retardance maxima were then binned by raft radius and the average value of each bin was plotted. Subsequently, the measurements were repeated for different background membrane compositions to determine the dependence of raft tilt on different background and raft chiralities. The low retardance of long-rod rafts made particle tracking from the retardance images impossible. The LC-PolScope imaging system calculates the local sample retardance from five images with different liquid crystal polarizer settings, including one in which the polarizer settings minimize the image intensity (*33*). In this image, a weak fluorescence signal is visible for the labeled long rod rafts. We tracked the rafts in these max extinction images and measured the retardance at those positions in each calculated image.

We used Python and OpenCV to analyze the dependence of the long-rod domain shape on domain area. We set a binary intensity threshold to distinguish the fluorescently labeled long-rod domain from the membrane background. We found the contour of the domain in the binary image and calculated the area within the contour. Using OpenCV's "minAreaRect" function to find the minimum bounding rectangle for each maximum contour, we measured the short and long sides of the bounding rectangle as the minor and major axes of the domain. To quantify the degree of circularity of each merging domain, we calculated the ratio of the minor axis to the major axis. A ratio of one indicates a square bounding rectangle and thus a circular shape, while a smaller ratio indicates an elongated shape. The main contour area and minor axis to major axis ratio was calculated for each image. We then plotted the ratios as a function of the areas of the long-rod domains and binned the data to see the resulting trends more clearly.

**Figures:**

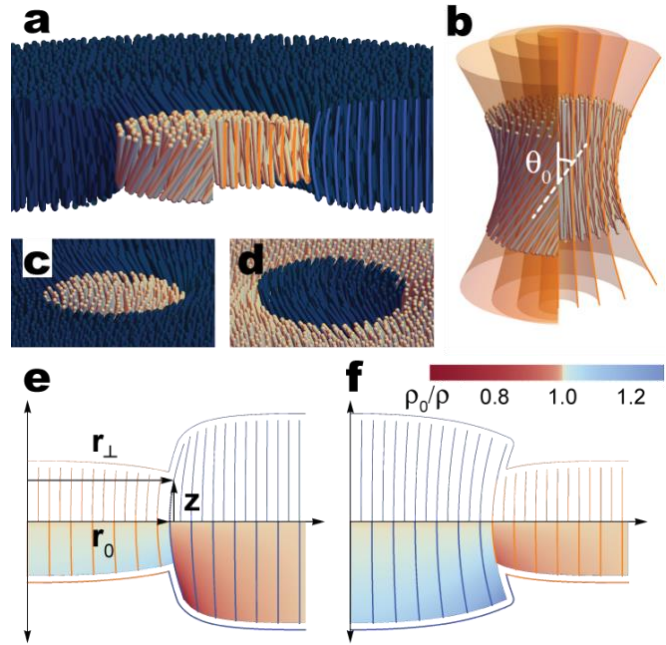

**Figure 1: Twist-splay coupling in colloidal rafts. a)** A twisted raft within a background membrane of longer rods. b) Hyperbolic structure of a raft composed of twisted rigid rods. Rods passing through the midplane at the same radial distance from the center define a hyperboloid surface. $\theta_0$ is the angle of maximum tilt with respect to vertical. **c, d)** Rafts composed of short and longer rods, respectively. **e)** Twist induces dilation of the short rods within the raft and compression of the long rods in the outer membrane, decreasing the total membrane volume that is inaccessible to depletants. Color indicates normalized local density ranging from low (blue) to high (red). **f)** Long-rod rafts have dilation of the long rods and compression of the short rods, leading to a net increase in the excluded volume and unfavorable structures.

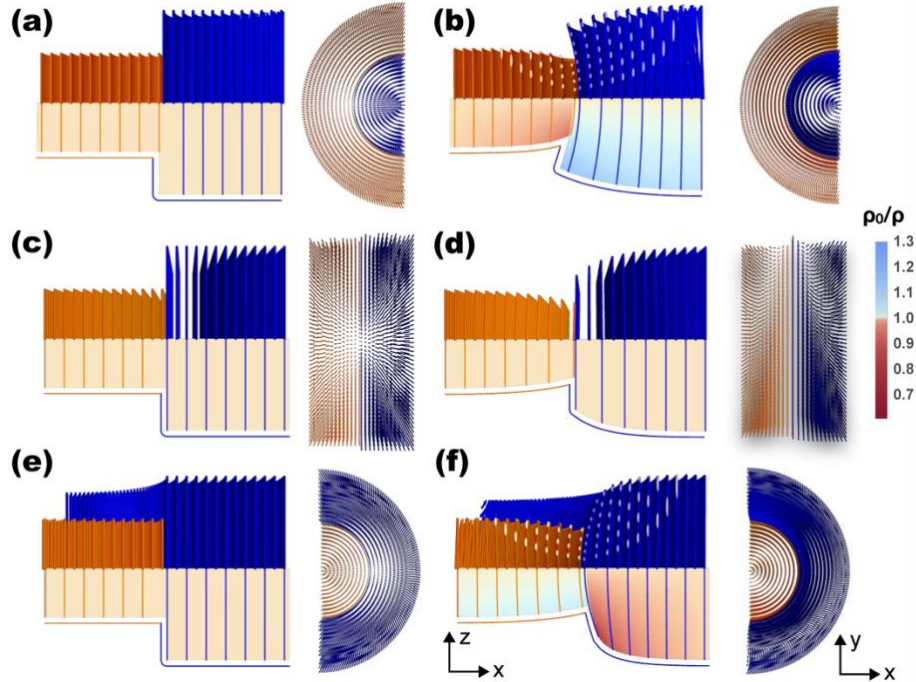

**Figure 2: Coupling between interface curvature rod tilt and density. a,b)** Structure of domain interface with curvature toward long rods ($\kappa_g > 0$). (left) Side and top views show the constant rod density of an untwisted curved domain. (right) The same interface with rod twist shows a splayed structure and expansion of the long rods and compression of the short rods. **c,d)** For interface without spontaneous curvature ($\kappa_g = 0$) the rod density is constant whether rods are straight or tilted. **e,f)** When $\kappa_g < 0$, rod density around the untwisted interface again remains constant, while rod twist leads to expansion of the short and compression of the long rods.

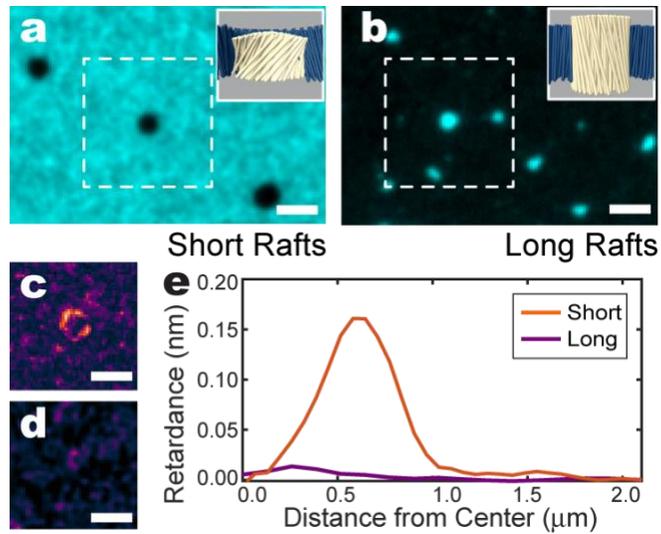

**Figure 3: Structure of short- and long-rod rafts. a)** Chiral short-rod rafts in a long-rod membrane of opposite chirality. Long rods are fluorescently labeled. Inset: schematic of raft structure. **b)** Rafts composed of fluorescently labeled long rods in a short-rod background membrane. **c)** A LC-PolScope image highlights the tilt localized at the edge of the short-rod the raft. Image intensity corresponds to the magnitude of the local tilt. **d)** A LC-PolScope image of a comparable long rod raft shows no measurable lilt. **e)** Radially-averaged retardance, which is proportional to the edge tilt, for the short- and long-rod rafts. Scale bars, 2 μm.

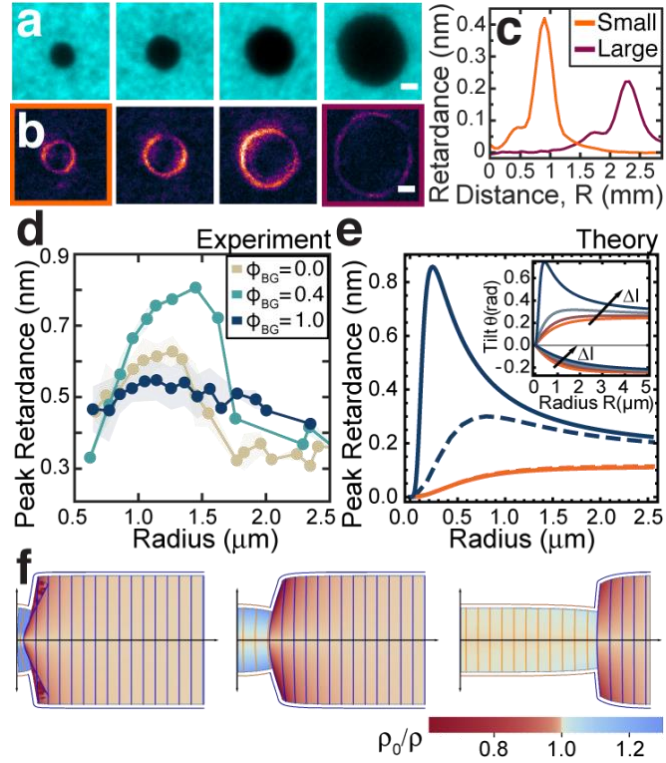

**Figure 4: Non-monotonic dependence of the edge twist on the raft size. a,b**) Fluorescence and LC-PolScope images of rafts of increasing sizes. Scale bars, 1 μm. **c**) Radially-averaged retardance for the smallest and largest rafts shown in b. **d**) Edge retardance plotted as a function of raft radius demonstrates non-monotonic dependence of twist on the raft size. Plots for short-rod rafts dissolved in background membranes with three different chiralities. $\phi_{BG} = 0$ (37% M13KO7 63% M13KO7-Y21M) indicates an achiral background membrane, while $\phi_{BG} = 1.0$ (100% M13KO7) indicates the strongest left-handed chirality. **e**) Predicted maximum retardance for Δl=0 (orange line) and Δl=0.26 μm (blue line). Solid lines represent linear model predictions, while dashed lines represent predictions of the full non-linear calculation. Inset: Predicted maximum raft tilt for length differences of Δl = ±0.1, 0.2, 0.26 μm. Increasing Δl leads to a pronounced peak in the maximum retardance as a function of raft radius, while negative values of Δl predict the suppression of edge twist for long-rod rafts. **f**) Local membrane density maps with maximum raft twist held constant for three different raft radii, $r$ = 0.15, 1.0, 1.8 μm and Δl = 0.86 μm, illustrating the geometric origin of the non-monotonic stiffening/softening resistance to edge twist with size

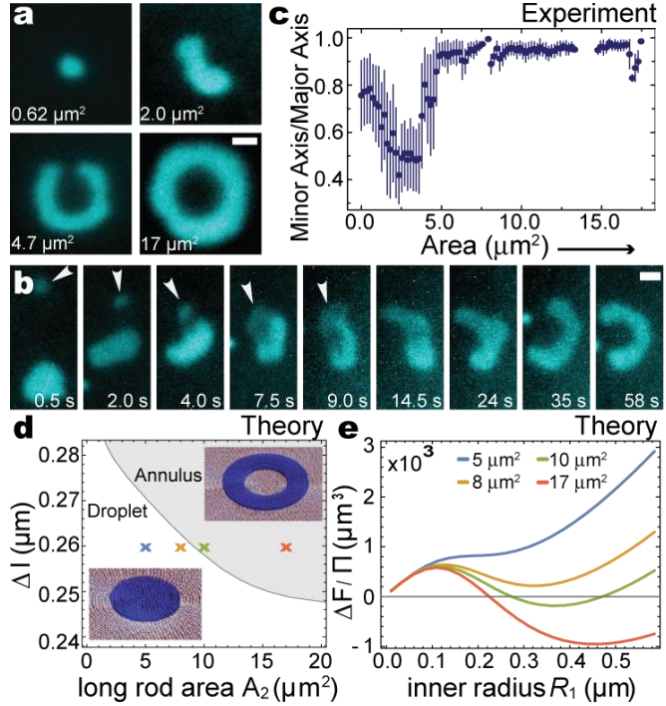

**Figure 5: Long rods form annular rafts. a**) Increasing the area of a long-rod raft induces a shape transition from a circular droplet to elongated shapes, before transforming into a horseshoe and eventually a complete annulus. **b**) An optical trap is used to fuse two long-rod droplets and the subsequent shape transformation into horseshoe shape. Arrow indicates the trap position. **c**) The ratio of minor to major axis as a function of the raft area. Error bars represent standard deviation of the average minor to major axis ratio for each area bin. **d**) Theoretical prediction for the phase diagram of long-rod raft shapes as a function of raft area and the length difference between two rod species. A length difference of 0.26 μm corresponds to the experimental parameters. **e**) Long-raft free energy landscape as a function of its size indicates the of long raft domains across droplet radii/annulus inner radii. Line colors correspond to the marked locations in phase diagram. The minima show the preferred droplet or annulus size. Scale bars, 1 μm.

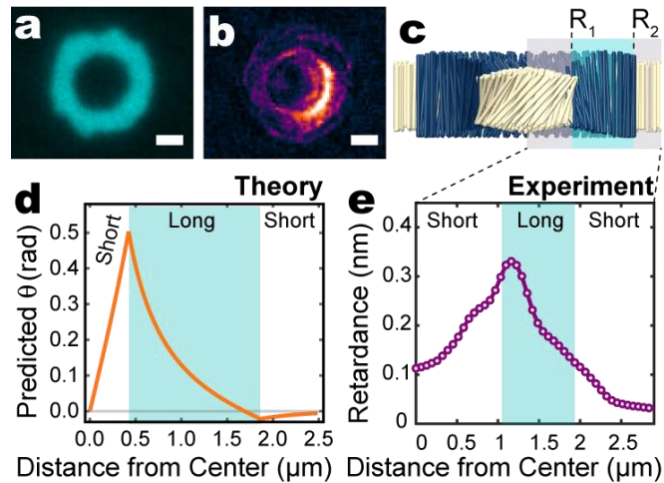

**Figure 6: Structure of annular long-rod rafts. a, b**) Fluorescence and LC-PolScope image of annular long-rod raft. **c**) Schematic of long-rod annulus. **d**) Theoretical predictions for the radial rod twist exhibit a maximum at the inner edge, and weak twist in the opposite direction at the outer edge. The annular width is highlighted in cyan. **(e)** Radially averaged retardance shows a peak in retardance at the inner edge which decays into the membrane background. The annulus width as measured using fluorescence is highlighted in cyan. Scale bars are 1 μm.

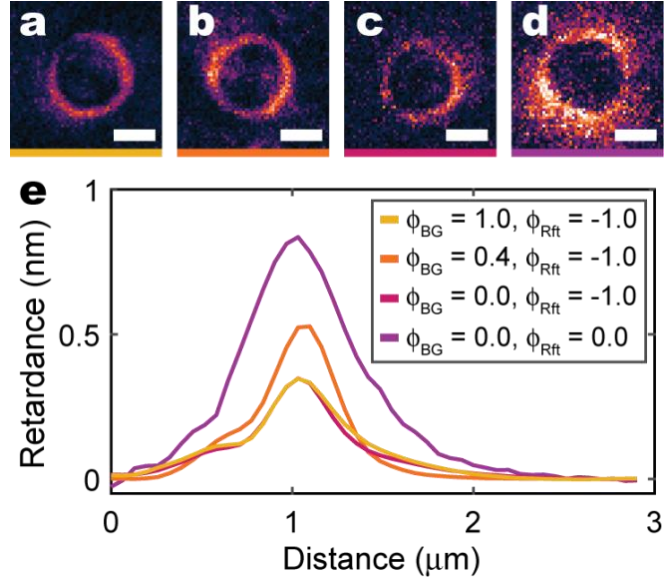

**Figure 7: Short achiral rafts in long achiral membranes are stable and exhibit edge twist.** Spontaneous twist of rafts composed of different mixtures of *fd-wt* and *fd-Y21M* and different compositions of the background membrane assembled from M13KO7 and M13KO7-Y21M. **a-c)** LC-PolScope images of three different rafts assembled in a right-handed membrane background ($\phi_{RFT} = -1.0$). The three images from left to right correspond to decreasing left-handed chirality of the short rod rafts ($\phi_{BG} = 1.0$, $\phi_{BG} = 0.4$, $\phi_{BG} = 0.0$). **d)** Stable colloidal rafts exhibits spontaneous twist in the absence of chiral asymmetry ($\phi_{BG} = 0.0$, $\phi_{RFT} = 0.0$). **e)** Radially averaged retardance as a function of the distance from the raft center for the four rafts show above. Scale bars, 1 µm.

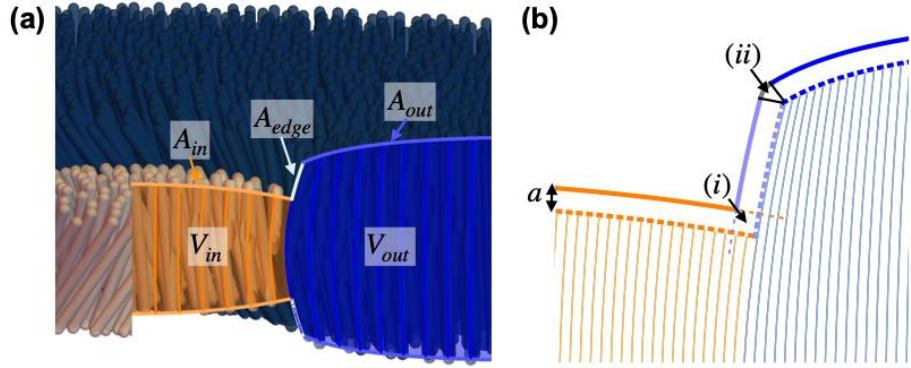

**Figure 8: Excluded volume of a twisted colloidal raft. a)** The volume excluded to the depletants is $V_{ex} = V_{in} + V_{out} + a\,(A_{in} + A_{edge} + A_{out})$ to first order in the depletant size. **b)** The nonlinear model excluded volume is found from the envelope of rod positions, shifted out by a distance of the depletant radius $a$. The resulting excluded volume "halo" finds the self-intersection of the shifted envelope at **(i)** to avoid double-counting the volume and smoothly joins the sections of the envelope at **(ii)** with a circular cap. Both contributions are second order in $a$.